\documentclass[amstex]{article}
\usepackage{amsmath}
\def\bp{{\mathbf p}}
\def\bx{{\mathbf x}}
\def\cK{{\mathcal K}}
\def\B{\mathcal{B}}
\def\Dir{\,\,{\raise.15ex\hbox{/}\mkern-12mu D}}
\def\a{\alpha}

\def\d{\delta}
\def\D{\Delta}
\def\g{\gamma}
\def\G{\Gamma}
\def\l{\lambda}
\def\m{\mu}
\def\n{\nu}
\def\vf{\varphi}
\def\h{\chi}
\def\r{\rho}

\def\pa{\partial}

\def\({\left(}
\def\){\right)}
\def\[{\left[}
\def\]{\right]}

\begin{document}
\ \\

\begin{center}
{\Large{\bf Note on the boundary terms in AdS/CFT correspondence
for Rarita-Schwinger field}}\\

\ \\

{\bf R.C.Rashkov}\footnote{
e-mail: rash@phys.uni-sofia.bg}  \\
\ \\
Department of Theoretical Physics\\
Sofia University, 5 J.Bourchier Blvd.\\
1164 Sofia, Bulgaria

\ \\

\end{center}

\ \\

\begin{abstract}
In this letter the boundary problem for massless and massive
Rarita-Schwinger field in the AdS/CFT correspondence is considered.
The considerations
are along the lines of a paper by Henneaux \cite{H} and are based on the
requirement the solutions to be a stationary point for the action
functional. It is shown that this requirement, along with a definite
asymptotic behavior of the solutions, fixes the boundary term that must
be added to the initial Rarita-Schwinger action. It is also shown  that
the boundary term reproduce the known two point correlation functions of
certain local operators in CFT living on the boundary.
\end{abstract}

\section{Introduction}

Recently a fascinating conjecture by Maldacena \cite{Malda} has been
proposed. According to \cite{Malda} the
supergravity theory in d+1 dimensional Anti-de Sitter space (AdS) with a
compact extra space is related by holographic correspondence principle
\cite{Sus} to a certain conformal field theory (CFT) living on the boundary
of AdS space. The underlying principle behind this AdS/CFT
correspondence was elaborated in explicit form by Gubser, Klebanov and
Polyakov \cite{gkp} and Witten \cite{w}. According to \cite{gkp} and
\cite{w}, the action for the supergravity theory on AdS considered as a
functional of the asymptotic values of the fields on the boundary is
interpreted as a generating functional for the correlation functions in the
conformal field theory living on the boundary. The explicit form of this
interpretation is:
\begin{equation}
\int\limits_{\Phi}{\mathcal D}\Phi exp\{-S\[\Phi\]\}=
\langle exp\int\limits_{\pa AdS}d^d{\mathbf x}{\mathcal O}\Phi_0\rangle
\end{equation}
where $\Phi_0$ is the boundary data for AdS theory which couples to a
certain conformal operator $\mathcal O$ on the boundary. This interpretation
has already a large number of examinations by computing various correlation
functions  of a local operators in CFT induced by AdS scalar fields
\cite{w,4,5,6,7}, spinor fields \cite{HS,8}, vector fields \cite{w,6,8,9},
antisymmetric fields \cite{14,15,16} and Rarita-Schwinger fields
\cite{C,A,kr}. All the examples confirmed the validity of the AdS/CFT
correspondence principle. The essence of these examinations is in
studying of the field behavior near the boundary of AdS space and
calculation of the boundary terms which produce the corresponding
correlation functions in CFT. While in the case of an action of AdS theory
with derivatives of order higher than one the considerations are, more or less,
transparent, in the case of Dirac-like actions (Dirac and Rarita-Schwinger)
the situation is more subtle. That is because the naive limit to the boundary
lead to vanishing of the action (it is zero on shell) which obviously
spoil the correspondence principle.

Recently two different but equivalent ways of treatment of the boundary term
problem in the case of spinor field were proposed \cite{ar,H}. In \cite{ar}
the considerations are based on the Hamiltonian approach treating $x^0$ as
an evolution parameter. A half of the components of the boundary spinor
turn out to be cannonically conjugated to the other half and the
boundary term naturally appears to be $p\dot q $.

In the paper \cite{H} another approach is used, namely the stationary phase
method. It is based on the fact that one can expand the path integral of a
given theory with action $S[\Phi]$:
$$
{\mathbf Z}=\int{\mathcal D}\Phi e^{\frac{i}{\hbar} S[\Phi]}
$$
near the stationary point given by the solutions of the classical theory. In
the classical limit ($h\to 0$) the leading order is simply:
$$
{\mathbf Z}\sim exp\{\frac{i}{\hbar} S_{class}\}
$$
where $S_{class}$ is the action functional evaluated on the classical
field configurations. The stationary point is determined by the requirement:
\begin{equation}
\frac{\d S}{\d\Phi}=0
\label{a}
\end{equation}
which is nothing but  the classical equations of motion with
given boundary data. The main immediate but important observation is that while
the action $S[\Phi]$ may satisfy (\ref{a}) on the clasical field
configurations, it is not true in general if there is a surface term
$\B_\infty$, i.e. there may be $\d\B_\infty\neq 0$ and therefore
$\d(S+\B_\infty)\neq 0$. Such a situation appears for instance in
the gauge field theory where $\B_\infty$ gives the conserved charges of the
symmetry currents. It follows that in order the classical configurations to
be a true stationary point it is necessary a boundary term to be added
ensuring the wanted requirement. This scheme was used in \cite{H} in order
to obtain the boundary term in the case of spinor field $\Psi$.

In this letter we would like to study the boundary term in the case of
massless and massive Rarita-Schwunger field. In our analysis we will use
the receipt described  in \cite{H} for derivation of the boundary term in
the spinor case.

The paper is organized as follows. In Section 2 the derivation of the
boundary term in case of massless Rarita-Schwinger field is given. It is
shown that the result reproduce exactly the two-point correlation functions
derived in \cite{A}. In Section 3 the same analysis is performed in
case of massive Rrita-Schwinger field. The result for the correlation
functions is in complete agreement with that given in \cite{kr}. In the
Conclusions we give some remarks and brief comments.

%%%%%%%%%%%%%%%%%%%%%%%%%%%%%%%%%%%%%%%%%%%%%%%%%%%%%%%%%%%%
\section{Boundary term for massless Rarita-Schwinger field}

In this Section we will consider a massles Rarita-Schwinger field on
$AdS_{d+1}$ given by the following action \cite{N,A}:
\begin{equation}
S_{R-S}=\int\limits_{AdS}d^{d+1}x\sqrt{G}\bar\Psi_\m\(\G^{\m\n\l}
D_\n-m\G^{\m\n}\)\Psi_\l
\label{1}
\end{equation}
We used the notations:
\begin{align}
&\G^{\m\n\l}=\G^{\[\m\right.}\G^\n\G^{\left.\l\]},\quad
\G^{\m\n}=\G^{\[\m\right.}\G^{\left.\n\]}\\ \notag
\label{1'}&\Psi_{\m}=\Psi_{\m \a}
%\label{1'}
\end{align}
where $\G^\m$ are the gamma matrices in AdS connected to the flat gamma
matrices by the relation $\G^\m=e^\m_a\g^a$. The vielbein is given below
and the flat gamma matrices satisfy the usual anticommutation relations
$\{\g^a,\g^b\}=2\d^{ab}$. We choose to work in coordinates
$x^a=(x^0,x^i)=(x^0,{\mathbf x}); i=1,\dots d$ defining $d+1$-dimensional
Euclidean Anti-de Sitter space as Lobachevski upper half plane $x^0>0$ with a
metric of the form:
\begin{equation}
ds^2=\frac{1}{x_0^2}\(dx^0+d{\mathbf x}^2\)
\label{2}
\end{equation}
With this choice the vielbein and the corresponding non-zero components
of the spin connection are given by the expressions:
\begin{equation}
e^\m_a=\frac{\d^a_\m}{x^0};\quad \omega_i^{0j}=-\omega_i^{j0}=
\frac{\d^j_i}{x^0};\quad a=0,\dots d
\label{3}
\end{equation}
The boundary of the AdS space consists in a hypersurface $x^0=0$ and a
single point $x^0=\infty$.

In this frame the covariant derivative and the Dirac operator reads off:
\begin{equation}
D_\n=\pa_\n+\frac{1}{2x^0}\g_{0\n};\quad \G^\m D_\m=\Dir=x_0\g^0\pa_0+
x_0{\mathbf\g}.{\mathbf\nabla}-\frac d2\g^0
\label{4}
\end{equation}
where ${\mathbf\g}=(\g^i);\,\, {\mathbf\nabla}=(\pa_i);\,\, i=1\dots d$.

The equation of motion for Rarita Schwinger field following from the
action (\ref{1}):
\begin{equation}
\G^{\m\n\l}D_\n\Psi_\l=m\G^{\m\n}\Psi_\l
\label{5}
\end{equation}
can be rewitten in the form \cite{A}:
\begin{align}
&x_0\g^\n\pa_\n\psi_0-\(\frac d2+1\)\g_0\psi_0 +m\psi_0=0\\ \notag
\label{6}&x_0\g^\n\pa_\n\psi_i-\frac d2\g_0\psi_i+m\psi_i=\g_a\psi_0
%\label{6}
\end{align}
where $\psi_a=e^\m_a\Psi_\m$.

In order to find the boundary contributions we are interested in studying
of the behavior of the solutions near the boundary $x^0=0$. For this purpose
one can use the Frobenius procedure looking for solutions of the form \cite{H}:
$$
\(x^0\)^\r\sum\limits_{n=0}^\infty c_n^a\(\vec{\mathbf x}\)\(x^0\)^n
$$
Subsitution of these series into the equations of motion determines the
values of the parameter $\r$:
$$
\r=\frac d2\pm m+\d_{a0}
$$
(the values of $\r$ turn out to be the same as in the spinor case
\cite{H} for $a\neq0$). Therefore, we are dealing with two types of
solutions:
\begin{equation}
\psi_a^-\(x_0,\bx\)=\(x_0\)^{\frac d2-m+\d_{a0}}\vf_a\(\bx\)+
o\(\(x_0\)^{\frac d2-m+\d_{a0}}\)
\label{7}
\end{equation}
\begin{equation}
\psi_a^+\(x_0,\bx\)=\(x_0\)^{\frac d2+m+\d_{a0}}\h_a\(\bx\)+
o\(\(x_0\)^{\frac d2+m+\d_{a0}}\)
\label{8}
\end{equation}
where:
\begin{align}
&\g^0\vf_a\({\bx}\)=-\vf\({\bx}\)\notag\\
\label{9}&\g^0\h_a\({\bx}\)=\h\({\bx}\)
\end{align}
It follows that $\psi^{\pm}_a$ are eigenvectors of $1/2(I\pm\g_0)$ with
eigenvalues $\pm 1$ respectively. The conjugated Rarita-Schwinger field
can be treated analogously giving the following result:
\begin{align}
\label{10}&\bar\psi^+_a\(\bx\)=\bar\vf_a\(\bx\)\(x_0\)^{\frac d2+m+\d_{a0}}+
o\(\(x_0\)^{\frac d2+m+\d_{a0}}\)\\
\label{11}&\bar\psi^-_a\(\bx\)=\bar\h_a\(\bx\)\(x_0\)^{\frac d2-m+\d_{a0}}+
o\(\(x_0\)^{\frac d2-m+\d_{a0}}\)
\end{align}
where the fields $\bar\vf_a$ and $\bar\h_a$ are subject to the
constraints:
$$
\bar\vf_a\g^0=\bar\vf_a;\quad \bar\h_a\g^0=-\bar\h_a
$$
Note that using the equations of motion one can find the subleading
terms recursively but since they doesnt contribute to the boundary we
will skip their explicit form.

Let us consider the general solutions of the equations of motion:
$$
\psi_a\(x_0,]bx\)=\int d^d\bp e^{i\bp.\bx}\[F_+\(x_0,\bp\)\tilde\vf_a^+\(\bp\)
+F_-\(x_0,\bp\)\tilde\vf_a^-\(\bp\)\]
$$
Since the solutions must be regular in the bulk (up to
$x_0=\infty$) the boundary spinor fields $\tilde\vf^\pm_a$ are not independent
and the solutions can be expressed in terms of $\tilde\vf^-_a$ only
\footnote{In what follows we will always suppose that the Fourier transform
is well defined, i.e. $\vf_a\(\bx\)$ and $\bar\vf_a\(\bx\)$ are of
compact support and vanish at $\bx\to\infty$ and therefore can be
Fourier transformed.}\cite{A}:
\begin{equation}
\psi_0\(x_0,\bx\)=\int d^dpe^{i\bp
.\bx}\(x_0p\)^{\frac{d+3}{2}}\[i\frac{\hat p}{p}\cK_{m+\frac 12}\(x_0p\)+
\cK_{m-\frac 12}\(x_0p\)\]\tilde\vf^-_0\(\bp\)
\label{12}
\end{equation}
\begin{multline}
\psi_i\(x_0,\bx\)=\int d^dp\(x_0p\)^{\frac{d+1}{2}}
\{\[i\frac{\hat p}{p}\cK_{m+\frac 12}\(x_0p\)+
\cK_{m-\frac 12}\(x_0p\)\]\tilde\vf^-_i\(\bp\)\\
 +\[\(\(2m+1\)\frac{p_i\hat p}{p^2}-ip_ix_0+\g_i\)\cK_{m+\frac 12}
\(x_0p\)-\frac{p_i\hat p}{p}x_0\cK_{m-\frac 12}\(x_0p\)\]\tilde\vf^-_0
\(\bp\)\}
\label{13}
\end{multline}
($\hat p=\g^ip_i$).In order to see how the fields $\tilde\vf^-_a$ are
related to the fields $\vf^-_a$ (which gives the asymptotic of $\psi_a$
at $x_0\to 0$) we use the small argument expansion of the modified
Bessel function $\cK_\n$:
\begin{equation}
\cK_\n\(z\)=\frac 12\[\(\frac z2\)^{-\n}\G(\n)\[1+\dots\]+\(\frac
z2\)^\n\G(-\n)\[1+\dots\]\]
\label{14}
\end{equation}
where dots stands for positive integer powers of $z^2$. Subsitution of
(\ref{14}) into (\ref{12},\ref{13}) gives the following behavior:
\begin{align}
&\psi_0^+\(x_0,\bx\)=x_0^{\frac d2-m+1}\int d^dpe^{i\bp .\bx}\[i\hat p
p^{\frac d2-m}2^{m-\frac 12}\G\(m+\frac 12\)\]\tilde\vf^-_0\(\bp\)+
o\(\(x_0\)^{\frac d2-m+1}\)\notag \\
\label{15}&\psi_0^-\(x_0,\bx\)=x_0^{\frac d2+m+1}\int
d^dpe^{i\bp .\bx}\[\frac{
p^{\frac d2+m+1}\G\(\frac 12-m\)}{2^{m+\frac 12}}\]\tilde\vf^-_0\(\bp\)+
o\(\(x_0\)^{\frac d2+m+1}\)
\end{align}
\begin{multline}
\psi^-_i\(x_0,\bx\)=
x_0^{\frac d2-m}\int d^dpe^{i\bp .\bx}\{\[i\hat p
p^{\frac d2-m-1}2^{m-\frac 12}\G\(m+\frac 12\)\]\tilde\vf^-_i\(\bp\)
\\
+\[p^{\frac d2-m}2^{m-\frac 12}\G\(m+\frac 12\)\(\(2m+1\)\frac{p_i\hat
p}{p^2}+\g_i\)\]\}\tilde\vf^-_0\(\bp\)+o\(\(x_0\)^{\frac d2-m}\)
\label{16}
\end{multline}
\begin{equation}
\psi^+_i\(x_0,\bx\)=x_0^{\frac d2+m}\int d^dpe^{i\bp .\bx}\frac{p^{\frac
d2+m}\G\(\frac 12-m\)}{2^{m+\frac 12}}\tilde\vf^-_i\(\bp\)+
o\(\(x_0\)^{\frac d2+m}\)
\label{16'}
\end{equation}
The components of the Rarita-Schwinger field are subject to one more
constraint:
$$
\g^0\psi_0+\g^i\psi_i=0
$$
which relate the $\tilde\vf_0^-$ and $\tilde\vf_i^-$ components as
follows \cite{A}:
\begin{equation}
\tilde\vf_0^-\(\bp\)=-\frac{2ip_j\tilde\vf_j^-\(\bp\)}{\(2m+d-1\)p};
\quad \g^i\tilde\vf_i^-\(\bp\)=0
\label{19}
\end{equation}
Using the asymptotic expressions for $\psi_a$ (\ref{7},\ref{8}) it is
easy to find the relation between $\vf_a\(\bp\)$ and $\h_a\(\bp\)$ and
$\tilde\vf^-_a\(\bp\)$:
\begin{align}
&\vf_0\(\bp\)=i\hat pp^{\frac d2-m}2^{m-\frac 12}\G\(\frac 12+m\)
\tilde\vf_0^-\(\bp\)
\notag\\
\label{17}&\h_0\(\bp\)=\frac{p^{\frac d2+m+1}}{2^{m+\frac 12}}\G\(\frac 12-m\)
\tilde\vf^-_0\(\bp\)
%\label{17}
\end{align}
and:
\begin{align}
&\vf_i\(\bp\)=p^{\frac d2-m}2^{m-\frac 12}\G\(m+\frac 12\)\[i\frac{\hat
p}{p}\tilde\vf_i^-+\(\(2m+1\)\frac{p_i\hat p}{p^2}+\g_i\)\tilde\vf^-_0\]
\notag\\
\label{18}& \h_i\(\bp\)=\frac{p^{\frac d2+m}\G\(\frac 12-m\)}{2^{m+\frac 12}}
\tilde\vf^-_i\(\bp\)
%\label{18}
\end{align}

From (\ref{15},\ref{16}, \ref{19}) it follows that $\psi^\pm_a$ can be
expressed in terms of $\vf^+_a\(\bp\)$ only and that $\h_a$ and $\vf_a$
are related "on-shell".

The final lesson from the above considerations is that a half of the
boundary data is expressible in terms of the other half but the
relations are valid "on-shell". The main conclusion is that the general
solutions of the equations of motion in the whole AdS space are
determined by the fields annihilated by $(I+\g_0)$, a quite similar
result as in case of spinor field \cite{H}.

Now we are going to apply the variational principle to the
Rarira-Schwinger action (\ref{1}). Since the Rarira-Schwinger equations
of motion are first order differential equations we cannot fix all the
components of $\psi_a$ at the boundary but only a half of them, $\vf_a$ or
$\h_a$. The basic idea is to use AdS correspondence principle which
tells us that the fields $\vf_a$ serve as a sources for the
bulk-boundary Green functions \cite{HS,C,A}. Thus, it is appropriate to fix
$\vf_a$ at the boundary and to leave $\h_a$ to vary. The variational
principle will be applied to all configurations of the form:
\begin{align}
&\psi_a=\psi^-_a+\psi^+_a\notag\\
\label{20}&\bar\psi_a=\bar\psi^-_a+\bar\psi^+_a
\end{align}
The fields $\psi^-_a$ and $\bar\psi^+_a$ have the asymptotic (\ref{7},
\ref{10}) while $\vf_a$ and $\bar\vf_a$ are fixed on the boundary. The
other part of $\psi^+_a$ and $\bar\psi^-_a$ behaves near the boundary as
it is described in (\ref{8},\ref{11}), but in this case the values of
$\h_a$ and $\bar\h_a$ on the boundary are free to vary. The relations between
$\vf_a$ and $\h_a$ (\ref{17}, \ref{18}) are only on-shell and will not
affect the variarional principle (the same is true for $\bar\vf_a$ and
$\bar\h_a$).

After the above preparations we are ready to vary the action (\ref{1})
with respect to $\psi_a$ and $\bar\psi_a$. As in \cite{H}, the variation
will be in the class of fields defined in (\ref{20}) but varying (\ref{1})
we will take into account the surface terms:
\begin{equation}
\d S_{R-S}=B_\infty+\[\,\,0\,\,\]_{on-shell}
\label{21}
\end{equation}
where:
\begin{multline}
B_\infty=-\frac 12\int d^d\bx\[\bar\vf_i\(\bx\)g^{ij}\d\h_j\(\bx\)+
\d\bar\h_i\(\bx\)g^{ij}\vf_j\(\bx\)+
\bar\vf_i\(\bx\)\g^i\g^j\d\h_j\(\bx\)\right.\\
\left.+\d\bar\h_i\(\bx\)\g^i\g^j\vf_j\(\bx\)\]
\label{22}
\end{multline}
(we recall that $\sqrt{G}=\(x_0\)^{-d-1}$ and $g^{ij}$ is the induced
metric on the boundary).

The term $B_\infty$ is nothing but the variation of the surface term at
infinity \cite{H}:
\begin{equation}
B_\infty=-\d C_\infty
\label{23}
\end{equation}
where:
\begin{equation}
C_\infty=\frac 12\int d^d\bx\[\bar\vf_ig^{ij}\h_j+\bar\h_ig^{ij}\vf_j+
\bar\vf_i\g^i\g^j\h_j+\bar\h_i\g^i\g^j\vf_j\]
\label{24}
\end{equation}
Note that since $\g^i\h_i=0$ the last two terms don't contribute. The
requirement for the action $S_{R-S}$ to be stationary on the solutions of
the equations of motion imposes to consider a new, improved action of
the form:
\begin{equation}
S=S_{R-S}+C_\infty
\label{25}
\end{equation}
It is obvious that $\d S=0$ on-shell and reproduce the correct solutions
of the equations of motion.

Of course, the above boundary term is not unique. This can be achieved
by imposing three natural conditions, namely:

a) $C_\infty$ is local

b) $B_\infty$ is without derivatives

c) $C_\infty$ preserves the AdS symmetry.

Under the above requirements $C_\infty$ is unique.

Having the explicit solutions for $\psi_a$ and its asymptotics, one can
rewrite the boundary term (up to irrelevant for our considerations
contact terms) as:
\begin{equation}
C_\infty=\lim\limits_{\varepsilon\to 0}\frac 12\int\limits_{M_\varepsilon}d^d
\sqrt{g_\varepsilon}\bar\psi_ig^{ij}\psi_j
\label{26}
\end{equation}
where $M_\varepsilon$ is a d-dimensional surface approaching the
boundary $\varepsilon\to 0$ and $g_\varepsilon$ in the induced metric on
$M_\varepsilon$. The boundary term (\ref{26}) is in complete agreement
with that of \cite{C,A}.

Using the explicit expressions for ($\vf_i,\h_i$) and
($\bar\vf_i,\bar\h_i$) it is straightforward to calculate the correlation
functions produced on the boundary. Since the Rarita-Schwinger action is
zero on-shell the contributions will come only from the boundary terms
(\ref{24}). According to the AdS/CFT correspondence principle one must replace
$\h_i$ and $\bar\h_i$ in (\ref{24}) with their on-shell values
(\ref{18}). The substitution gives:
\begin{multline}
S_{class}=\int\frac{d^d\bp}{(2\pi)^d}\[\bar\vf_i\(-\bp\)\h_i\(\bp\)+
\bar\h_i\(\bp\)\vf_i\(-\bp\)\]\\
=i\frac{\G\(\frac 12-m\)}{2^{2m}\G\(\frac
12+m\)}\int\frac{d^d\bp}{(2\pi)^d}\bar\vf_i\(\bp\)\(\d_{ij}\frac{\hat
p}{p}-\frac{2\(2m+1\)}{d+2m-1}\frac{p_ip_j\hat p}{p^3}\)\vf_i\(\bp\)\\
=-\frac{\G\(\frac{2m+d+1}{2}\)}{\pi^{\frac d2}\G\(\frac 12+m\)}\int
d^d\bx d^d{\mathbf y}\bar\vf_i\(\bx\)\frac{\(x-y\)_i\g^i}{|\bx-{\mathbf
y}|^{2m+d+1}}\[\d_{ij}-2\frac{\(x-y\)_i\(x-y\)_j}{|\bx-{\mathbf y}|^2}\]
\vf_j\({\mathbf y}\)
\label{27}
\end{multline}
The above expression coincides with the two point correlation functions
found in \cite{C,A}:
\begin{equation}
\Omega\(\bx,{\mathbf y}\)\sim\frac{\(x-y\)_i\g^i}{|\bx-{\mathbf
y}|^{2m+d+1}}\[\d_{ij}-2\frac{\(x-y\)_i\(x-y\)_j}{|\bx-{\mathbf y}|^2}\]
\label{28}
\end{equation}
corresponding to conformal operator of dimension $\D=\frac d2+m$.
%%%%%%%%%%%%%%%%%%%%%%%%%%%%%%%%%%%%%%%%%%%%%%%%%%%%
\section{Massive case}

We proceed with the analysis of the boundary term in the case of massive
Rarita-Schwinger field. The most general action is given by \cite{N}:
\begin{equation}
S_{mR-S}=\int d^d\bx\sqrt{G}\[\bar\Psi_\m\G^{\n\n\r}D_\n\Psi_\r-
-m_1\bar\Psi_\m\Psi^\n-m_2\bar\Psi_\m\G^{\m\n}\Psi_\n\]
\label{30}
\end{equation}
The equations of motion following from (\ref{30})
$$
\G^{\m\n\r}D_\n\Psi_\r-m_1\Psi^\m-m_2\G^{\m\n}\Psi_\n=0
$$
can be rewritten in more convenient form \cite{C,kr}:
\begin{equation}
\G^\n\(D_\n\Psi_\r-D_\r\Psi_\n\)-m_-\Psi_\r+\frac{m_+}{d-1}\G_\r
\G^\n\Psi_\n=0
\label{31}
\end{equation}
where we use the notations of \cite{kr}: $m_\pm=m_1\pm m_2$. Applying
the standard procedure of passing to flat space equations ($\psi_a=
e^\m_a\Psi_\m$) it is straightforward to obtain the equations for
$\psi_0$ and $\psi_i$ \cite{C,kr}:
\begin{align}
\label{32}&x_0\g^a\pa_a\psi_0-\(\frac d2+1\)\g_0\psi_0-m_-\psi_0=
x_0\pa_0\eta-
\frac{m_+}{d-1}\g_0\eta-\eta\\
\label{33}&x_0\g^a\pa_a\psi_i-\frac d2\psi_i-m_-\psi_i=
x_0\pa_i\eta+\frac 12\g_0
\g_i\eta-\frac{m_+}{d-1}\g_i\eta+\g_i\psi_0
%\label
\end{align}
where $\eta=\g^a\psi_a$.

One can try to apply the standard Frobenius procedure in solving the
system (\ref{32},\ref{33}) by looking for solutions in form:
\begin{equation}
\psi^a\sim\(x_0\)^\r\sum\limits_{n=0}^\infty c_n^a\(\bx\)x_0^n
\label{35}
\end{equation}
where $c_n^a(\bx)$ are $x_0$-independent Rarita-Schwinger fields. If one
try to solve the system for fields depending on $x_0$ only the leading
terms will have rather different values for $\r$ \footnote{As it was
noted in \cite{kr}  simple
algebraic operations leads to the following relation between $\psi_0$ and
$\eta$ when we consider the dependence on $x_0$ {\it only}:
$$
\(d-1-2m_-\g_0\)\psi_0=\(d-1+2m_2\g_0\)\eta.
$$
Note that $b^\pm_a$ are constant spinors.} \cite{kr}:
\begin{align}
\label{36}&\psi_0\sim x_0^{\frac d2+C}b_0^++x_0^{\frac d2-C}b_0^-\\
\label{37}&\psi_i\sim x_0^{\frac d2+C}\g_ib_0^++x_0^{\frac d2-C}\g_ib_0^-
+x_0^{\frac d2+m_-}b_i^++x_0^{\frac d2-m_-}b_i^-
%\label{36,37}
\end{align}
where the constant $C$ is given by:
\begin{equation}
C=\frac{d\(d-1\)}{4m_1}+\frac{m_-\(m_1+dm_2\)}{m_1\(d-1\)}
\label{38}
\end{equation}
Similar expressions for the conjugated fields hold. The splitting of
$\vf_a$ into $\pm$ parts is subject to the conditions:
\begin{equation}
\g_0b_a^\pm=\pm b^\pm_a;\quad \bar b_a^\pm=\mp\bar b^\pm_a
\label{41}
\end{equation}

Let us discuss the general solutions of the system (\ref{32}, \ref{33}).
For the $\psi_0$ component it reads off\footnote{In order to introduce $\bx$
dependence, in \cite{kr} O(d+1,1) transformations are used. Such
transformations are accompanied by rotation for the spinor fields. One can
show that our expressions are equivalent to those of \cite{kr}.} \cite{kr}:
\begin{multline}
\psi_0\(x_0,\bp\)=
\left\{\frac{x_0^{\frac{d+1}{2}}p^{\frac 12-C\g_0}}{2^{\frac 32-C\g_0}
\G\(\frac d2+\frac 32-C\g_0\)}\[\(d-3+2C\g_0+2i\hat px_0\)\cK_{\frac 12
-C\g_0}\(x_0p\)\right.\right.\\
+\(2x_0\g_0p-i\(d+3-2C\g_0\)\frac{\hat p}{p}\)\cK_{\frac
12+C\g_0}\(x_0p\)-\\
\left.\frac{4m_1x_0p}{d\(d-1-2m_2\)}\(\(\frac{d}{x_0}-i\g_0\hat p\)\cK_{\frac
12-C\g_0}\(x_0p\)-p\cK_{\frac 12+C\g_0}\(x_0p\)\)\]\tilde\vf_0\(\bp\)\\
\left. -\frac{2x_0^{\frac{d+1}{2}}p^{\frac 12-m_-\g_0}}{2^{\frac
32-m_-\g_0}\G\(\frac d2+\frac 32-m_-\g_0\)}\(-i\g_0p_i\cK_{\frac
12-m_-\g_0}\(x_0p\)-\frac{p_i\hat p}{p}\cK_{\frac 12+m_-\g_0}\(x_0p\)\)
\tilde\vf_i\right\}
\label{42}
\end{multline}
\begin{multline}
\psi_i\(x_0,\bp\)=
\left\{\frac{x_0^{\frac{d+1}{2}}p^{\frac 12-C\g_0}}{2^{\frac 32-C\g_0}
\G\(\frac d2+\frac 32-C\g_0\)}\[2\frac{p_i\hat p}{p}x_0\cK_{\frac 12+
C\g_0}\(x_0p\)\right.\right.\\
-2\(\g_i-ip_i\g_0x_0\)\cK_{\frac 12-C\g_0}\(x_0p\)
+\frac{2m_1}{d\(d-1+2m_2\)}\times\\
\times\(\(\(d-1-2C\g_0\)\g_0\g_i+2ip_ix_0\)
\cK_{\frac 12-C\g_0}\(x_0p\)
-\(\(d-1-2C\g_0\)i\frac{\hat p}{p}\g_i\right.\right.\\
\left.\left.\left.+2i\(1+2C\g_0\)\frac{p_i}{p}-
-2i\frac{dp_i+\g_i\hat p}{p}
-2\frac{p_i\hat p}{p}\g_0x_0\)\cK_{\frac
12+C\g_0}\(x_0p\)\)\]\tilde\vf_0\(\bp\)\\
+\frac{x_0^{\frac{d+1}{2}}p^{\frac 12-m_-\g_0}}
{2^{\frac 32-m_-\g_0}\G\(
\frac d2+\frac 32-m_-\g_0\)}
\[\(\(d-1-2m_-\g_0\)\g_0\d_{ij}+2i\frac
{p_ip_j\hat p}{p^2}x_0\)\cK_{\frac 12-m_-\g_0}\(x_0p\) \right.\\
-\(\(d-1-2m_-\g_0\)i\frac{\hat p}{p}\d_{ij}+2i\(1+2m_-\g_0\)\frac
{p_ip_j\hat p}{p^3}-2i\frac{p_j\g_i}{p}\right.\\
-\left.\left.\left. \frac{2p_ip_j}{p}\g_0 x_0\)
\cK_{\frac 12+m_-\g_0}\]\tilde\vf_j\(\bp\)\right\}
\label{43}
\end{multline}
where $\tilde\vf_a$ satisfy the relations:
%\begin{align}
$$
\g_0\tilde\vf_a=-\tilde\vf_a;\quad
\g^i\tilde\vf_i=0\notag
%\end{align}
$$
and analogous expression for conjugated fields. Since we have two rather
different leading terms (of powers $d/2\pm C$ and $d/2\pm m_-$),
some of the solutions have to be fixed to zero, i.e. we must fix
$\tilde\vf_0$ or $\tilde\vf_i$.
A natural criterion for this is the requirement that in
the limit $m_2\to 0$ to reproduce the massless case. This uniquelly
determine $\tilde\vf_0=0$. The same arguments as in the massless case
for regularity of the solution in
the interior of the AdS relate $\tilde\vf_i^+$ and $\tilde\vf_i^-$
which reminiscent again
the principle that only a half of the components can be fixed on the boundary.
\footnote{This requirement lead to the condition $p_j\tilde\vf_j^+=
-i\frac{\(d-1+2m_-\)}{\(d-1-2m_-\)}p^{2m_-}\frac{\hat p}{p}p_j\tilde\vf_j^-$.
The relation holds only on-shell. Note that in our expressions the spinors
are rotated compared to \cite{kr} and will be denodet by $\hat{}$.}
In order to extract the contribution to the boundary we will use the small
argument expansion of the modified Bessel function (\ref{14}). The result
for $\psi_0$ is:
\begin{align}
&\psi_0\(x_0,\bp\)=x_0^{\frac d2+m_-+1}\h_0\(\bp\)+
        x_0^{\frac d2-m_-+1}\vf_0\(\bp\)\label{45}\\
&\h_0=\frac{\G\(\frac 12-m_-\)}{d-1+2m_-}
          \[
          \frac{p_i\hat p}{p}\(\frac p2\)^{2m_-}\hat\vf_i^- \]\label{46}\\
&\vf_0=-\frac{\G\(\frac 12+m_-\)}{d-1+2m_-}
          \[ip_i\hat\vf^-_i\]\label{47}
\end{align}
One can express $\h_0$ in terms of $\vf_0$, but since $\psi_0\sim
x_0^{d/2\pm m_-+1}$ there will be no contribution to the boundary term.

Let us apply the same analysis to the $\psi_i$ components. We split $\psi_i$
into "chiral"  and "anti-chiral" parts:
\begin{align}
\label{48}&\psi_i\(x_0,\bp\)=
x_0^{\frac d2+m_-}\h_i\(x_0,\bp\)+x_0^{\frac d2-m_-}\vf_i\(x_0,\bp\)\\
\label{49}&\g_0\h_i=\h_i;\quad \g_0\psi_i=-\psi_i
\end{align}
The analysis of the $x_0\to0$ behavior of the corresponding component gives
the following expressions:
\begin{align}
&\vf_i\(\bp\)=\frac{\G\(\frac 12-m_-\)}{2^{m_-+\frac 12}}\(
\(d-1+2m_-\)i\frac{\hat p}{p}\d_{ij}+2i\(1-2m_-\)\frac{p_ip_j\hat p}
{p^3}-2i\frac{p_j\g_i}{p}\)p^{2m_-}\hat\vf_j^-\label{50}\\
&\h_i\(\bp\)=-\frac{\G\(\frac 12+m_-\)}{2^{-m_-+\frac 12}}\(d-1+2m_-\)
\hat\vf_i^-\label{51}
\end{align}
From the above formulae follows that the two expressions (\ref{50},
\ref{51}) are not independent and a half of the boundary data can be
expressed (on-shell) in terms of the other half.

We now turn to the variational principle applied to the action for the
massive Rarita-Schwinger field (\ref{30}). Repeating the same considerations
as in the massless case we have found the boundary terms in the same form
\footnote{We note that since $\g^i\h_i=0$ the $3^{th}$ and $4^{th}$ terms
in (\ref{55}) and the second term in (\ref{57}) below
will not contribute.}
:
\begin{equation}
C_\infty=\frac 12\int d^d\[\bar\vf_ig^{ij}\h_j+\bar\h_ig^{ij}\vf_j+\bar\vf_i
\g^i\g^j\h_j+\bar\h_i\g^i\g^j\vf_j\]
\label{55}
\end{equation}
which again can be written as in the massless case:
\begin{equation}
C_\infty=\lim\limits_{\varepsilon\to 0}\frac 12\int\limits_{M_\varepsilon}d^d
\sqrt{g_\varepsilon}\(\bar\psi_ig^{ij}\psi_j+\bar\psi_i\g^i\g^j\psi_j\)
\label{57}
\end{equation}
The action ensuring that the classical solutions are true stationary point
of the action (\ref{30}) gets the modification:
\begin{equation}
S=S_{mR-S}+C_\infty
\label{56}
\end{equation}
which is unique under the requirement of locality, absence of derivatives
and presevation of the AdS symmetry.

Now, using (\ref{50}, \ref{51}) it is straightfoward to reproduce the
correlation functions in the CFT living on the boundary found in \cite{kr}:
\begin{equation}
\Omega\(\bx,{\mathbf y}\)\sim\frac{\(x-y\)_i\g^i}{|\bx-{\mathbf
y}|^{2m_-+d+1}}\[\d_{ij}-2\frac{\(x-y\)_i\(x-y\)_j}{|\bx-{\mathbf y}|^2}\]
\label{58}
\end{equation}
Note that in the massless limit $m_2\to 0$ the above result coincide with
(\ref{28}).

%%%%%%%%%%%%%%%%%%%%%%%%%%%%%%%%%%%%%%%%%%%%%%%%%%%%

\ \\

{\bf{\large Conclusions}}
\ \\
In this letter we have analysed the boundary term for Rarita-Schwinger
field in the AdS/CFT correspondence. It is shown that, as in the spinor case,
one cannot fix simultaneously all the components of the Rarita-Schwinger
field on the boundary but only a half of them. Following \cite{H} we used the
stationary phase method to determine the surface term.
We apply variation of the action over
an appropriate class of field configurations. Since we are dealing with
a theory with a boundary, it turns out that one half of the components of the
spinor field must be kept fixed but the other half are free to vary on the
boundary (at $x_0=0$). We choose to impose the boundary conditions on the
"chiral" part of the Rarita-Schwinger field (annihilated by $1/2(I-\g_0)$)
which have a minimal value of the leading term in $x_0$. The choice is quite
natural thinking of this part as of a source of the bulk-boundary Green
function. The "anti-chiral" starts at higher power of $x_0$ but it is shown
that these components play an important role since they contribute to the
boundary term. Moreover, these components are not off-shell subject to
boundary conditions.

It would be interesting to study interacting Rarita-Scwinger--scalars in
AdS/CFT correspondence as in \cite{Ge} and to proceed with an
investigation of the $S$-matrix along the lines of \cite{Sus1,P,Bala,Ar}.
Work on the subject is in progress.

\ \\

{\bf Note added}

After this paper was completed I was informed about paper \cite{af}
where, in the case of $AdS_5\times S^5$ geometry of type IIB
supergravity, analogous boundary terms for massless gravitino field were found.
I thank Dr. S.Frolov for information and comments. In \cite{vl} was
described a procedure of obtaining the boundary terms based on the
intertwining operators considerations. I am obligated to Prof. V.K.
Dobrev for information.

\ \\

{\bf Acknowledgments}

I am grateful to N.I.Karchev and M.Stanishkov for comments and critical
reading the manuscript.

\ \\

\end{document}